# Meta-analysis of Gene Expression in Neurodegenerative Diseases Reveals Patterns in GABA Synthesis and Heat Stress Pathways

Abdulahad Bayraktar

Bioinformatics Department, Graduate School of Natural and Applied Sciences, Mugla Sitki Kocman University, Kotekli, Mugla Turkey, abdulahadbayraktar@gmail.com

Regenerative and Restorative Medicine Research Center, REMER, İstanbul Medipol University, Istanbul, Turkey, abdulahadbayraktar@medipol.edu.tr

Tuğba Önal-Süzek

Bioinformatics Department, Graduate School of Natural and Applied Sciences, Mugla Sitki Kocman University, Kotekli, Mugla Turkey, tugbasuzek@mu.edu.tr

Barış Ethem Süzek

Bioinformatics Department, Graduate School of Natural and Applied Sciences, Mugla Sitki Kocman University, Kotekli, Mugla Turkey, barissuzek@mu.edu.tr

Ömür Baysal

Molecular Biology and Genetics Department, Graduate School of Natural and Applied Sciences, Mugla Sitki Kocman University, Kotekli, Mugla Turkey, omurbaysal@mu.edu.tr

## 1 Introduction

Neurodegenerative diseases are defined as loss of structure and function of specific neural cell groups resulting in mortality. These diseases are mostly associated with aging, due to the high prevalence in elder age groups (Feigin et al., 2017). Although aging association has been shown experimentally to a degree (Kritsilis et al., 2018; Reeve, Simcox, & Turnbull, 2014), aging is not the sole reason and many factors, such as familial background, stressful lifestyle and viral infection, are considered having causative effects (Fargo & Bleiler, 2014; Feigin et al., 2017).

For several neurodegenerative diseases, genetic factors are identified. A well-known example is the case of Huntington's disease. It was shown that an allele of the huntingtin gene (HTT) which has an extensive number of CAG repeats syntheses a form of huntingtin having polyglutamine repeats. This polymorphic form of the protein is folded abnormally; hence causing aggregation of the protein in cytosol and nucleus (Difiglia et al., 2016; Kaltenbach et al., 2007). These aggregates interact with a diverse range of proteins (Li & Li, 2006), cause apoptosis of neurons (Liu, 1998) and drive the pathogenesis of Huntington's disease (Landles & Bates, 2004). Similar findings can be presented for other neurodegenerative diseases, such as ALS and Alzheimer Disease (Kiernan et al., 2011; Kolarova, García-Sierra, Bartos, Ricny, & Ripova, 2012; Selkoe, Hardy, Selkoe, & Hardy, 2016). Although there are strong theories about the pathogenesis of the diseases, clinical trials of treatment by targeting molecules set by these theories did not become successful overall. Proposed treatments, e.g. vaccines (Schneeberger et al., 2009), are mostly palliative (Fargo & Bleiler, 2014; Jankovic, 2008); many could not be able to pass beyond phase 3 level trials (Doody et al., 2013; Karran & Hardy, 2014).

The pathophysiologic overlap between different neurodegenerative diseases is widely studied. Neurodegenerative diseases share common biological processes (e.g. protein aggregation, glial activation), chemical alterations (e.g. increase in the levels of cytokines and reactive oxygen species), (Wood, Winslow, Strasser, Wood, & Israel, 2015) and gene associations (Goedert, Clavaguera, & Tolnay, 2010).

Advancement of high-throughput sequencing technologies, particularly RNA Sequencing (RNA-Seq) led to an increased availability of public data for several diseases including neurodegenerative diseases. This made it possible to conduct meta-analysis, co-interpretation of the results of multiple studies, on multiple and independent RNA-Seq studies to increase the statistical power of the results (Rau, Marot, & Jaffrézic, 2014; Sweeney, Haynes, Vallania, Ioannidis, & Khatri, 2017) and, even, identify novel findings. Indeed, a meta-analysis of RNA-Seq of different types of diseases or similar diseases can unveil common transcripts. For instance, a large-scale RNA-Seq study put forth common genes of clusters of 21 cancer types using samples from TCGA (Peng et al., 2015).

In our work, we carried out a meta-analysis of RNA-Seq studies for three neurodegenerative diseases, namely Alzheimer's disease, Parkinson's disease and Amyotrophic Lateral Sclerosis (ALS). Meta-analysis can work well on large datasets from diverse samples with multiple factors. An inspiring study was performed by Hamel Patel, Richard J.B Dobson, Stephen J Newhouse, is a good example of this (Patel et al., 2018). In this study, more than 2500 microarray datasets mapping to four brain regions samples from multiple AD, HD, schizophrenia, bipolar disorder, major depressive disorder, and PD studies were meta-analysed. Differential expression analysis was performed on each of them similar to our work. However, a different p-value combining method, which is more suitable for microarray data, were applied. By this meta-analysis, they determined AD specific significant differentially expressed genes (DEGs) in different brain regions. In another study, researchers focused on common features, i.e. genes and ontologies, of neurodegenerative disorders rather than differences. They studied 1270 post-mortem central nervous system tissue samples, microarray data, from 13 patient cohorts of four NDD, namely AD, HD, ALS and PD (M. D. Li, Burns, Morgan, & Khatri, 2014). In this particular study, however, they used an effect size combination method. Based on their findings, they suggested the impaired mitochondrial function as the most reliable specific feature of neurodegeneration.

In our paper, we describe our approach to RNA-Seq data collection and analysis, report on the gene expression, pathway and mutation load patterns for significantly differentially expressed genes (DEGs) in all three diseases.

## 2 DATA

We searched the Sequence Read Archive (SRA) (Leinonen, Sugawara, & Shumway, 2011) and European Nucleotide Archive (ERA) (Harrison et al., 2019) databases for RNA sequence data, with the terms "parkinson", "alzheimer" and "amyotrophic". We narrowed our findings to publicly available, human, post mortem, non-blood tissue studies. In the end, we identified 4 studies fitting with our selection criteria in the SRA database.

PRJNA279249 (Prudencio et al., 2015) and PRJNA306659 were identified for ALS. PRJNA279249 has 33 fALS and sALS and 17 control samples, taken from the cerebellum and prefrontal cortex sequenced by Illumina HiSeq 2000. The library used is paired and the average sequence length (forward + reverse) is 202. PRJNA306659 has 13 sALS and 8 control samples, sequenced by Illumina Genome Analyzer II. Motor neurons dissected from the spinal cord were studied. Library prepared was single, 50-nucleotide was the average sequence length.

PRJNA283498 (Dumitriu et al., 2016) was identified for PD. It has 44 PD and 29 control samples, sequenced by Illumina HiSeq 2000. Prefrontal cortex was used as sample. Library used is paired and average sequence length (forward + reverse) was 202. Due to low quality, one PD and one control sample were excluded from the PD accession table, SRR2015769, and SRR2015725, respectively. PRJNA377568 (Friedman et al., 2018) was identified for AD. It has 84 patients and 33 controls. Samples were dissected from fusiform gyrus, sequenced by Illumina HiSeq 2500. Library used was paired and average sequence length (forward + reverse) was 200.To study on them, accession tables including sample IDs and symbols of disease status were prepared for each study. Disease statuses provided in run information tables were accepted as it is except for PRJNA279249, in which we considered both C9orf72-associated fALS and sALS samples as ALS. The FASTQ files containing both forward and reverse pairs are split as separate files for forward and reverse reads for further analysis. In total, 259 samples (173 patients + 86 control) were studied.

## 3 METHOD

Our study is comprised of four stages: computing DEGs for the ALS, Parkinson's and Alzheimer's disease datasets, analysis of variants for all the datasets, and network analysis for the DEGs, and finally, comparison of the DEGs, pathways, and variants between these neurodegenerative datasets.

### 3.1 Differential Expression Analysis of the Parkinson's, ALS and Alzheimer's Disease Datasets

First, we checked the quality of samples using FastQC (Andrews, 2010) and all the samples met the default quality metrics. Non-coding RNAs were removed using a FASTA file composed of human ribosomal and transfer RNA sequences downloaded from NCBI RefSeq (O'Leary et al., 2016). Second, we aligned reads to reference genome

GRCh38.p12 (Schneider et al., 2017) using STAR (Dobin et al., 2013) alignment tool and annotated reads with corresponding GTF. In this part, sjdbOverhang parameter was set according to length of average length of pairs that were given in SRA, and twopassMode was set as Basic to reduce misalignments. In addition, *quantMode* parameter allowed us to count reads per gene. We verified on log files that the average of the uniquely mapped read frequency was higher than 90% for all reads for each data set. Last, we merged gene count tables into a single file for each study. These merged files were studied in the following step.

Using the DESeq2 package (Love, Huber, & Anders, 2014) of R on the gene count tables with default parameters, then DEGs between control and patient samples for all three diseases (4 datasets) were identified using FDR as p-value adjustment and setting %5 as p-value threshold. Next, the DEGs from four datasets were inner joined into a single table in Rstudio. Then, we performed a meta-analysis on the table using the metaRNASeq package (Rau et al., 2014). The success of p-value combination methods in meta-analysis has been confirmed in several similar comparative studies (Chang, Lin, Sibille, & Tseng, 2013; Marot, Foulley, Mayer, & Jaffrezic, 2009; Sweeney et al., 2017). metaRNASeq provides two p-value combination techniques, namely inverse normal and Fisher, for differential meta-analyses of RNA-Seq data. Inverse normal and Fisher functions operate on raw p-values and they apply Benjamini-Hochberg correction on p-values to control false discovery rate at 0.05 at default. We tried both methods on the q-values with a threshold of 0.05 and narrowed down our gene list to genes that are identified as differentially expressed (up or down-regulated) in the same direction in all four datasets as suggested by the metaRNASeq article.

## 3.2 Variant Analysis of the Parkinson's, ALS and Alzheimer's Disease Datasets

For variant analysis, we used the alignments computed in Section 3.1 and identified variants by rnaseqmut (https://github.com/davidliwei/rnaseqmut) core script with the parameters to select variants that do not exist in any control sample but exists in at least 2 patient samples. For each disease, disease-associated variants were filtered into another VCF (Danecek et al., 2011) using rnaseqmut scripts. Variants were annotated using ANNOVAR (Wang, Li, & Hakonarson, 2010). Annotated variant tables were merged into a single spreadsheet.

## 3.3 Pathway Analysis of the Parkinson's, ALS and Alzheimer's Disease Datasets

We explored the common Reactome (Fabregat et al., 2018) pathways for the DEGs from section 3.1 In Cytoscape (Paul Shannon et al., 2003), the potential relations/links between these genes are examined using ReactomeFI plugin (Wu, Feng, & Stein, 2010) and overrepresented Gene Ontology (Gene Ontology Consortium, 2004)(GO) terms using BiNGO plugin (Maere, Heymans, & Kuiper, 2005). We built a network of significantly overrepresented pathways concerning these genes and combined with Parkinson, Alzheimer and ALS pathways using Wikipathways plugin (Kutmon, Lotia, Evelo, & Pico, 2014) of Cytoscape. We mapped expression (up- vs. down-regulation) and number variant information on the created networks.

# 4 RESULTS

Differential expression analysis of four datasets that we carried out had 13603 genes in common. We obtained 1,365 and 2,954 significant genes for Fisher and inverse normal p-value combination methods respectively, whose Benjamini-Hochberg adjusted p-values were lower than 0.05. The genes categorized as significantly differentially expressed using both p-value combination methods and having the same regulation pattern in all four datasets (e.g., upregulated in all or downregulated in all when compared to controls) resulted in a final list of 200 genes. In R, we compared these genes with disease-associated gene lists which we downloaded from NCBI RefSeq. 12 of the 200 genes (CALN1, VGF, ADAMTSL1, FKBP4, PTGES3, AHSA1, HSP90AA1, GAS7, PKN1, C22orf34, SPR, and RBMS1) have known associations with Alzheimer's, Parkinson's and ALS diseases based on previous studies.
Reactome enrichment of these 200 genes revealed overrepresentation of pathways mainly related to cellular responses to stress (e.g. attenuation phase, HSF-1 dependent transactivation pathways, regulation of HSF1-mediated heat shock response, HSP90-chaperone cycle for steroid hormone receptors) and GABA synthesis (e.g. MECP2 regulates transcription of genes involved in GABA signaling) (Figure 1). Genes overrepresented in the top three pathways were HSP90AA1, PTGES3, FKBP4, DNAJB1, DNAJB6, DEDD2 and HSPB1.

**Figure 1: The overrepresented pathways in Reactome for the 200 significantly DEGs common for three diseases.**

GO biological process enrichment validates Reactome enrichment results and further explains the significance of some

| Pathway name | Entities | | | | Reactions | |
|---|---|---|---|---|---|---|
| | found | ratio | p-value | FDR* | found | ratio |
| Attenuation phase | 9 / 47 | 0.003 | 7.96e-08 | 4.12e-05 | 4 / 5 | 4.13e-04 |
| HSF1-dependent transactivation | 9 / 59 | 0.004 | 5.24e-07 | 1.36e-04 | 5 / 8 | 6.60e-04 |
| MECP2 regulates transcription of genes involved in GABA signaling | 4 / 6 | 4.24e-04 | 2.95e-06 | 5.08e-04 | 4 / 4 | 3.30e-04 |
| HSF1 activation | 7 / 43 | 0.003 | 6.47e-06 | 8.35e-04 | 4 / 7 | 5.78e-04 |
| TFAP2A acts as a transcriptional repressor during retinoic acid induced cell differentiation | 4 / 9 | 6.36e-04 | 1.44e-05 | 0.001 | 4 / 7 | 5.78e-04 |
| HSP90 chaperone cycle for steroid hormone receptors (SHR) | 8 / 70 | 0.005 | 1.81e-05 | 0.002 | 12 / 12 | 9.91e-04 |
| Transcriptional regulation by the AP-2 (TFAP2) family of transcription factors | 7 / 52 | 0.004 | 2.17e-05 | 0.002 | 9 / 44 | 0.004 |
| Cellular response to heat stress | 10 / 135 | 0.01 | 6.51e-05 | 0.004 | 13 / 29 | 0.002 |
| Regulation of HSF1-mediated heat shock response | 7 / 113 | 0.008 | 0.002 | 0.128 | 4 / 14 | 0.001 |
| Cellular responses to stress | 17 / 514 | 0.036 | 0.003 | 0.17 | 33 / 184 | 0.015 |
| GABA synthesis | 2 / 7 | 4.95e-04 | 0.006 | 0.262 | 2 / 2 | 1.65e-04 |

pathways. The genes associated with protein folding, refolding and responding to unfolded protein are also the ones represented in the cellular response to heat stress pathways, except DEDD2 and HSPB1. Besides, GAD1 and GAD2, as they are involved in glutamate decarboxylation to succinate, are represented in the GABA synthesis pathway (Table 1).

**Table 1: Overrepresented GO Biological Processes for 200 significant DEGs. Produced by BiNGO.**

| Description | p-value | Corrected p-value | Cluster frequency | Total frequency | Genes |
|---|---|---|---|---|---|
| Glutamate decarboxylation to succinate | 8.4556E-5 | 2.4487E-2 | 2 / 132 (1.5%) | 2 / 14301 (0.0%) | GAD1 GAD2 |
| Protein folding | 3.8739E-6 | 1.8698E-3 | 10 / 132 (7.5%) | 170 / 14301 (1.1%) | DNAJB2 DNAJA1 DNAJB1 HSP90AA1 DNAJB6 PTGES3 AHSA1 FKBP4 PPID HSPD1 |
| Protein refolding | 8.4307E-7 | 1.2208E-3 | 4 / 132 (3.0%) | 9 / 14301 (0.0%) | DNAJB1 HSP90AA1 PTGES3 HSPD1 |
| Response to protein stimulus | 1.2493E-5 | 4.5223E-3 | 8 / 132 (6.0%) | 117 / 14301 (0.8%) | DNAJB2 DNAJA1 DNAJB1 HSP90AA1 DNAJB6 HSPB1 BCL2L1 HSPD1 |
| Response to unfolded protein | 2.4088E-6 | 1.7440E-3 | 7 / 132 (5.3%) | 66 / 14301 (0.4%) | DNAJB2 DNAJA1 DNAJB1 HSP90AA1 DNAJB6 HSPB1 HSPD1 |

We performed gene set analysis by ReactomeFI without linker genes option using 200 genes list. In this way, genes were connected via pathway interactions pulled from Reactome. Figure 2 illustrates the components of the network produced. Red nodes represent upregulated genes, while blue nodes represent the downregulated genes. Interaction network presents that *HSP90AA1, PTGES3, FKBP4, AHSA1, DNAJA1, DNAJB1, PPID, and HSPD1* are connected and up-regulated in the largest connected component of the cluster. Downregulated *GAD1* and *GAD2* also make a pair. We imported "Cellular response to heat stress" on Cytoscape via Wikipathways and merged with our gene list. DJNAB1 and DEDD2 were the only significantly DEGs, both upregulated (Figure 2).

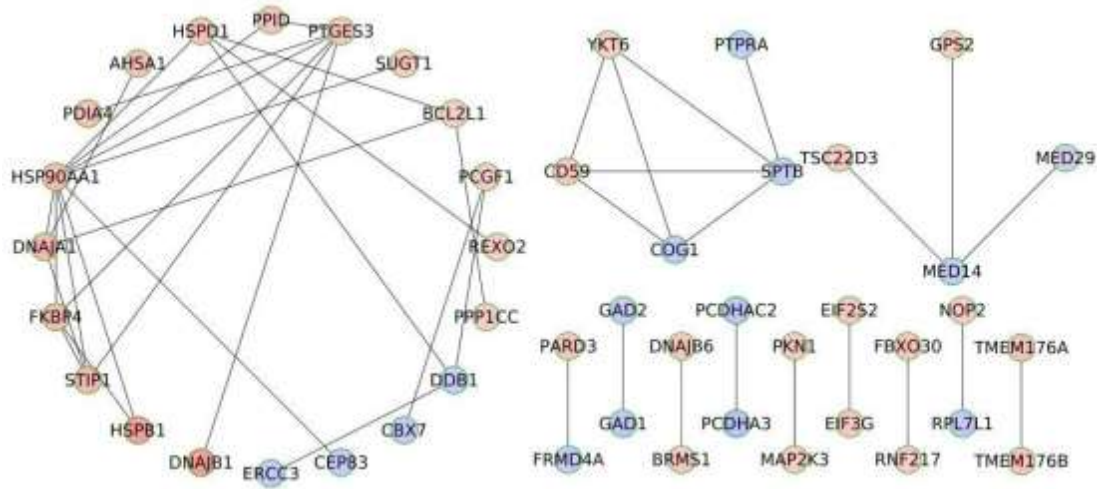

**Figure 2: Network of 47 found by ReactomeFI plugin based on 200 significant DEG.**

Finally, we compared the variants in all three diseases and found no variants common to all three diseases. Hence in the particular case, instead of intersecting the three diseases' variants, we merged the variant lists among which 3500 variants ended up with the SIFT predictor is D (deleterious). We filtered 850 nonsynonymous, exonic variants whose Polyphen2 HVAR predictor is not B (benign) from this list. Then, we compared the variant carrying genes to 200 common DEGs. We found 36 and 8 matches for genes in unfiltered and filtered variant lists, respectively. Interestingly, all detected variants of these genes had either highly low ($< 0.01$) or unknown frequencies overall considering 1000Genomes and ExAC. *GAD1, GAD2, HSP90AA1, HSPD1, BCL2L1, PPID* and *DNAJB6* are the 7 genes carrying mutations. These genes are primarily involved in cellular heat stress response and GABA synthesis pathways as mentioned above.

Overall, our findings overlap with the original studies to some degree. The AD and PD articles (Dumitriu et al., 2016; Friedman et al., 2018) shared differential expression analysis results. Of 200 genes of our meta-analysis, 117 and 45 were differentially expressed within FDR 0.05 cut-off of 0.05, respectively. Moreover, the ALS article (Prudencio et al., 2015) have lists of genes for significant GOs in the supplementary file, though it does not have a differential expression Table. HSP genes were confirmed by these original studies (PRJNA306659, PRJNA283498 and PRJNA377568) as differentially expressed within FDR 0.05 cut-off of 0.05. Besides, both *GAD1* and *GAD2* were significantly differentially expressed in AD but for PD only GAD2 was present, confirming our findings to an extent. The 4th study (ALS - PRJNA279249) does not have a publication so we could not compare.

## 5 DISCUSSION

In this study, we analyzed four independent datasets to explore their common DEGs and pathways. Based on our analysis, we identified the heat shock proteins and the related pathways' have a significant role in neurodegenerative diseases. Although we have not included Huntington's disease in our analysis, through a literature search we found out that the most significant shared pathway, attenuation phase, associated with heat shock proteins was also important for Huntington's disease (Kaltenbach et al., 2007). Moreover, heat shock proteins are recently in the radar for a potential target of neurodegenerative diseases (Kampinga et al, 2016). For instance, Colchicine, originally a medication to treat gout attacks, was recently repurposed as a potential vaccine for its function of enhancing the heat shock proteins on ALS patients (Crippa et al., 2016). (https://clinicaltrials.gov/ct2/show/NCT03693781 ).

In our analysis, we also identified two glutamic acid dehydrogenases GAD1 and GAD2, responsible for the synthesis of the GABA for synaptic release, which were both downregulated in all three of the diseases. This observation was supported by our discovery of variants in both GAD genes. Gamma-aminobutyric acid (GABA) is the primary inhibitory neurotransmitter in the cortex (Petroff, 2002). Decarboxylation of glutamate by glutamate decarboxylases, i.e. GAD1 and GAD2, is the crucial step in GABA synthesis (Bu et al., 1992). Unsurprisingly, a decrease in levels of GAD1 and GAD2 reduces GABA synthesis drastically, which in turn prevents the inhibition of postsynaptic GABAergic neurons (having GABA receptors). It can be speculated that this leads to continuous excitation of the postsynaptic neuron, or glutamate excitotoxicity, in other words. Through a literature search, we found that the loss of GABA receptors role in Alzheimer disease was well documented (Limon, Reyes-Ruiz, & Miledi, 2012). Furthermore, an ongoing clinical trial is testing whether the administration of flumazenil improves the motor function for the Parkinson disease patients with low GABA receptor availability (https://clinicaltrials.gov/ct2/show/NCT03462641). Hence, our study further supports the significance of GABA synthesis pathway and related proteins in neurodegenerative diseases.

In addition to DEG, we have explored the common mutation landscape of these three diseases. With a recent study showing the possibility of detecting somatic mutations in normal tissue using RNA-seq (Yizhak et al., n.d.)points to the potential of the increased mutation load of brain cells associated with neurodegenerative diseases. With the recently emerging importance of the somatic mutations in neurological disorders (Poduri, Evrony, Cai, & Walsh, 2013), the mutation dataset generated by our work has the potential to be used in further studies to find the set of somatic mutations in NDDs. Although these genes were reported for either one of these diseases separately in earlier studies, to our knowledge, no study so far reported their shared neurodegenerative role in all three diseases.

## ACKNOWLEDGMENTS

We would like to thank Medipol University Regenerative and Restorative Medicine Research Center for supporting this work by providing the high performance computing cluster on which we were able to perform this study's intensive time-consuming computations. We would like to also thank BAP 19/079/09/2/2 and BAP 19/079/10/2/2 projects for providing the travel support for the poster and the lightning talk presentations at the CNB-MAC 2019.